\documentclass[aps,PRApplied,reprint,superscriptaddress]{revtex4-2}
\usepackage{graphicx}

\begin{document}

%\preprint{}

\title{Quantitative measurement of the thermal contact resistance between a glass microsphere and a plate}

\author{Joris Doumouro}
\affiliation{Institut Langevin, ESPCI Paris, Université PSL, CNRS, 75005 Paris, France}
\author{Elodie Perros}
\affiliation{Institut Langevin, ESPCI Paris, Université PSL, CNRS, 75005 Paris, France}
\affiliation{Saint-Gobain Research Paris, 39 quai Lucien-Lefranc, Aubervilliers, France}
\author{Alix Dodu}
\affiliation{Institut Langevin, ESPCI Paris, Université PSL, CNRS, 75005 Paris, France}
\author{Nancy Rahbany}
\affiliation{Institut Langevin, ESPCI Paris, Université PSL, CNRS, 75005 Paris, France}
\author{Dominique Leprat}
\affiliation{LNE-Laboratoire national de m\'etrologie et d'essais, 78197 Trappes, France}
\author{Valentina Krachmalnicoff}
\affiliation{Institut Langevin, ESPCI Paris, Université PSL, CNRS, 75005 Paris, France}
\author{R\'emi Carminati}
\affiliation{Institut Langevin, ESPCI Paris, Université PSL, CNRS, 75005 Paris, France}
\author{Wilfrid Poirier}
\affiliation{LNE-Laboratoire national de m\'etrologie et d'essais, 78197 Trappes, France}
\author{Yannick De Wilde}
\email[]{yannick.dewilde@espci.fr}
\affiliation{Institut Langevin, ESPCI Paris, Université PSL, CNRS, 75005 Paris, France}

\date{\today}

\begin{abstract}
Accurate measurements of the thermal resistance between micro-objects made of insulating materials are complex because of their small size, low conductivity, and the presence of various ill-defined gaps. 
We address this issue using a modified scanning thermal microscope operating in vacuum and in air. The sphere-plate geometry is considered.  Under controlled heating power, we measure the temperature on top of a glass microsphere glued to the probe as it approaches a glass plate at room temperature with nanometer accuracy. In vacuum, a jump is observed at contact. From this jump in temperature and the modeling of the thermal resistance of a sphere, the sphere-plate contact resistance $ R_K=(1.4 \pm 0.18)\times10^7 \ \mathrm{K.W^{-1}}$ and effective radius $r=(36 \pm 4)$ nm are obtained. In air, the temperature on top of the sphere shows a decrease starting from a sphere-plate distance of 200 $\mathrm{\mu m}$. A jump is also observed at contact, with a reduced amplitude.  The sphere-plate coupling out of contact can be described by the resistance shape factor of a sphere in front of a plate in air, placed in a circuit involving a series and a parallel resistance that are determined by fitting the approach curve. The contact resistance in air $R^*_K=(1.2 \pm 0.46)\times 10^7 \ \mathrm{K.W^{-1}}$ is then estimated from the temperature jump. The method is quantitative without requiring any tedious multiple-scale numerical simulation, and is versatile to describe the coupling between micro-objects from large distances to contact in various environments. 

\end{abstract}

\maketitle

\section{\label{sec:Introduction}Introduction}
The effective thermal conductivity of complex insulating materials drastically depends on their micro-structure \cite{KWON_IJHMT2009,LANGLAIS_TI2004}.
In general, they are made of low thermal conductivity micrometer-sized elements in contact or near to contact. Enhanced insulation results from a high resistance between them that hampers heat flow through the material. It is of crucial importance to have a detailed knowledge of the thermal resistance between the constitutive micro-elements of insulating materials, both in air and in vacuum, but accurate measurements are complex to perform because of their small size and very low thermal conductivity. To describe the heat transfer processes at a microscopic scale, a distinction must be made between situations in which micro-elements are close to contact or in contact. Different heat transfer channels are expected to contribute between very close micro-elements. If they are separated by a vacuum gap smaller than the thermal wavelength $\lambda_{BB}\approx$ 10 $\mu$m at room temperature, the radiative heat transfer should be enhanced by the contribution of evanescent waves \cite{de_wilde_thermal_2006,kittel_near-field_2005,wischnath_near-field_2008,rousseau_radiative_2009,shen_surface_2009,ghashami_precision_2018,cui_study_2017}, but under ambient conditions this radiative contribution is outplayed by conductive and convective transfer. If the micro-elements are in contact, an additional heat transfer occurs due to conduction through the solid phase and a water meniscus potentially surrounding the contact area \cite{assy_temperature-dependent_2015}. 

In insulating materials made of fibers or grains, micro-elements are separated by a large variety of gaps. A well-known example is glass wool, a biphasic material made of randomly distributed glass fibers with a diameter on the order of tens of micrometers, interacting at various distances, from hundreds of micrometers to contact. From a theoretical point of view, handling this variety of scales is a challenge for numerical simulations. Therefore an analytical model would be versatile to estimate the heat flux as a function of the gap. Yet, there has been little effort to analytically model heat transfer between two bodies separated by small gaps or in contact under ambient conditions \cite{batchelor_thermal_1977}.

On the experimental side, scanning thermal microscopy (SThM) is ideally suited to perform local thermal investigations which require a control of the position of a small thermal sensor with nanometer accuracy. It consists of an atomic force microscope probe including a very small thermometer on its tip, whose position with respect to the sample surface is adjusted by means of piezoelectric translations. Such a sensor is generally employed for thermal characterization with nanometer spatial resolution of nanomaterials, nanostructures, and nanoscale devices \cite{gomes_scanning_2015,menges_temperature_2016}. It has also been used to investigate the thermal contact resistance between the probe tip and a planar substrate \cite{jarzembski_feedback_2018,lefevre_nanoscale_2006,park_experimental_2008,shi_thermal_2002,zhang_noncontact_2011} and to measure the thermal resistance due to a water meniscus formed around the probe-substrate contact region \cite{assy_heat_2015}. Most SThM studies use a thermal sensor placed in direct contact with the sample, which raises subtle metrological issues \cite{ramiandrisoa_dark_2017,gomes_scanning_2015}. In recent studies a microsphere was glued on the probe of a SThM to study the near-field radiative heat transfer in a sphere-plate geometry on a phase transition material \cite{menges_thermal_2016} or on a photovoltaic cell \cite{lucchesi_harnessing_nodate}.

In this paper we propose a SThM-based method to investigate the thermal resistance between a glass microsphere and a glass plate separated by a gap varying from hundreds of micrometers to contact,  both under vacuum and ambient conditions. The strategy relies on the use of a SThM probe working in active mode, i.e. both as thermometer and heater, with a glass microsphere glued on its tip. The transfer between the glass sphere and the plate is disentangled from the transfer between the heater and the environment through other paths, by monitoring the temperature at the top of the sphere as it approaches the plate and enters in contact with it. The study intends to provide a better insight regarding how heat is transferred between micrometer-sized glass elements within complex insulating materials like glass wool. 

First, we describe the experimental setup. Then, we solve the problem of heat conduction in vacuum based on measurements combined with an analytical expression of the thermal resistance of a sphere, which leads to an accurate estimate of the sphere-plate thermal contact resistance and of the contact effective diameter. Finally, the heat conduction problem in ambient conditions is also tackled theoretically using a phenomenological approach based on a shape factor which properly describes the approach curve of the sphere towards the plate. A good estimate of the sphere-plate contact resistance is also obtained in these more complex conditions. 

\section{\label{sec:setup}Experimental setup}
\begin{figure}
	\includegraphics[width=8.5cm]{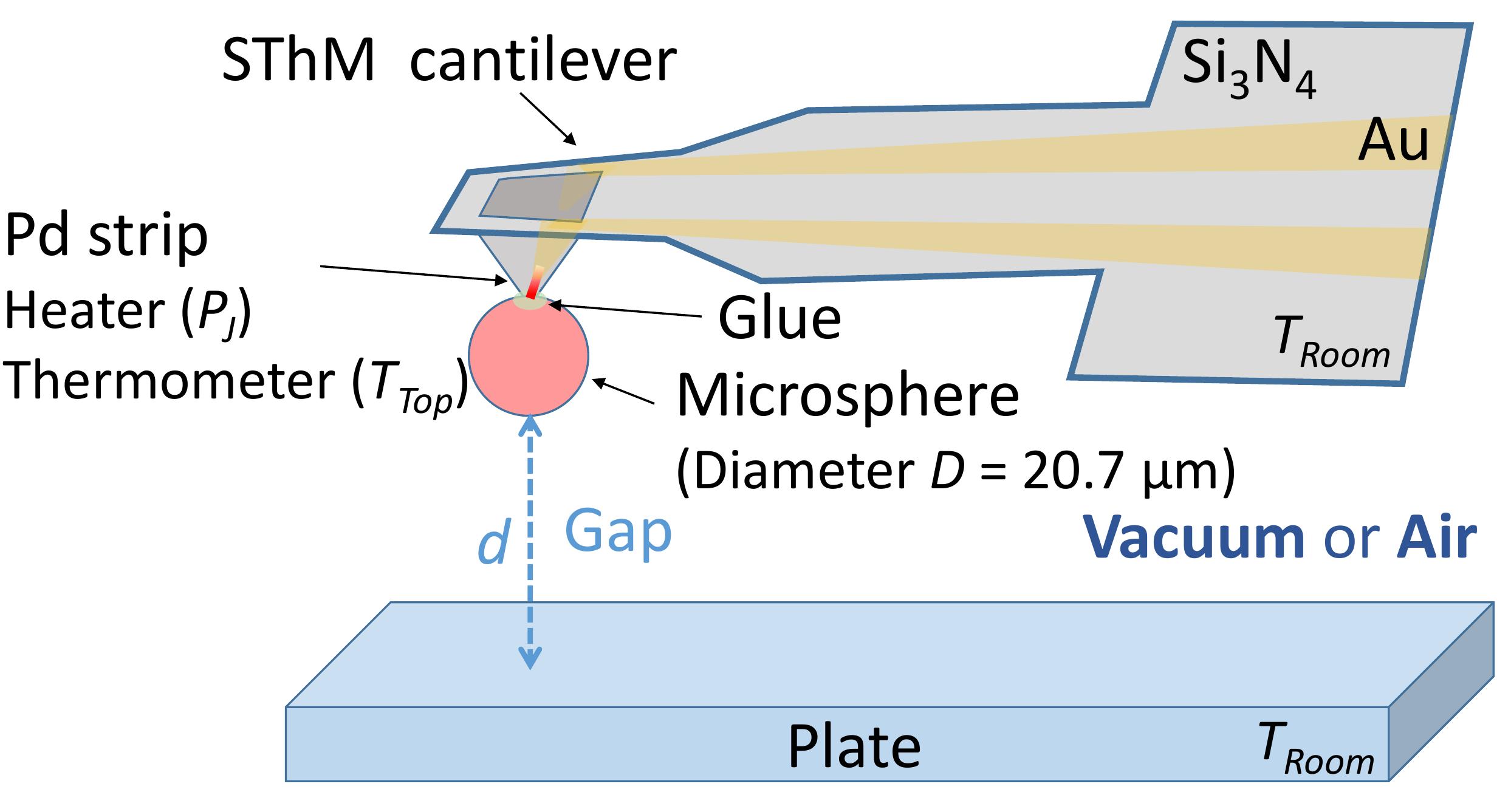}
	\caption{\label{figure_1}(a) Sketch of the SThM based setup used for the determination of the  contact or near-contact thermal resistance between a  microsphere and a plate made of borosilicate glass. A calibrated thermoresistive palladium (Pd) strip is used to heat the top of the microsphere by Joule heating using a DC current $I_{DC}$ while a small AC current $I_{AC}$ is applied to measure the temperature $T_{Top}$ via the strip resistance $\mathcal{R}(T)$ measured by means of a Wheatstone bridge, as described in the Appendix.}
\end{figure}

To investigate the sphere-plate geometry a borosilicate glass sphere of calibrated diameter $D=20.7 \pm 1 \mathrm{\mu m}$ from \emph{Duke Standards} is attached with epoxy glue to the truncated pyramidal tip of a SThM probe fabricated by \emph{Kelvin Nanotechnology}  (KNT), as shown in Fig.~\ref{figure_1}. The probe is made of silicon nitride (Si$_3$N$_4$) and carries a thermoresistive thin palladium strip patterned on the flat portion of the pyramidal tip. Its temperature coefficient of 0.74 $\Omega /K$ is obtained by placing the entire SThM probe in an isolated oven whose temperature can be accurately controlled while the strip electrical resistance $\mathcal{R}$ is measured. To perform the thermal measurements, we developed a setup which can operate either under vacuum or ambient conditions. It combines a \emph{SmarAct} linear piezoelectric stepper-motor and a \emph{Piezosystem Jena} linear continuous piezoelectric positionner. This allows us to perform displacements for measurements requiring a sphere-plate separation up to several hundreds of micrometers, and fine approach measurements when the sphere is close to contact with the plate. The latter is a borosilicate glass microscope coverslip. A secondary vacuum of approximately $10^{-6}$mbar is achieved by connecting a turbo pump to the chamber.
The top of the glass microsphere is glued to the palladium strip on the SThM probe. The latter is used both for local temperature measurements and heating.
The heated microsphere is held above the glass plate at a room temperature, $T_{Room}$. The gap between the base of the sphere and the plate surface is precisely adjusted with the piezoelectric stages while the temperature at the sphere top is monitored.

The SThM probes from KNT used in these experiments have a typical room temperature resistance of 350 $\Omega$. The temperature at the top of the microsphere $T_{Top}$ is obtained by measuring the electrical resistance of the thermoresistive palladium strip glued to it.
This electrical resistance $\mathcal{R}(T)$ is accurately measured by integrating it in a highly-sensitive Wheatstone resistance bridge built at LNE (see Appendix).  The bridge is biased with an AC voltage superimposed to a DC voltage which are controlled by the internal oscillator of a  lock-in detector and a \emph{Yokogawa} voltage source respectively. An AC polarization current $I_{AC}$ ($\approx 30 \ \mu$A at 873 Hz, typically) circulates through the palladium strip. The resistance $\mathcal{R}(T)$ is obtained from the three other reference resistances constituting the bridge and the AC voltage unbalance detected by the lock-in detector. $T_{Top}$ is then easily obtained based on a fine calibration of the temperature-resistance relationship carried out prior to measurements. We are able to resolve temperature variations on the order of 10 mK. To use the probe in active mode, a DC current $I_{DC}$ of a few hundreds ( $400 - 500 \ \mu$A, typically) is superimposed on the AC current. This produces a well-defined and well-controlled local Joule heating of the SThM tip and of the glass microsphere attached to it. This injected power, $P_J=\mathcal{R}{I_{DC}}^2$, was adjusted around 70 $\mathrm{\mu W}$. All measurements were performed without a deflection laser nor tip illumination, in order to prevent uncontrolled heat injection and to reduce measurement noise \cite{ramiandrisoa_dark_2017,spiece_improving_2018}.

In the experiments, the dissipated heat power $P_J$ is applied at the top of the glass microsphere and results in a temperature increase of the SThM probe above room temperature $T_{Room}$ ($\approx 30\mathrm{^\circ}$C in  the experiments discussed in this paper) whose magnitude  $\Delta T_{Top}=(T_{Top}-T_{Room})$ depends on the environmental conditions of the microsphere. This heat flux is dissipated through several channels including radiation between bodies, conduction in the air, the glass sphere, the SThM cantilever, through the sphere-plate contact, and convection. In vacuum, conduction and  convection in air are suppressed.

To characterize these thermal conductance channels, we measured the temperature $T_{Top}$ at the top of the sphere  as a function of the separation $d$ between its base and the glass plate from large distances to contact, first in vacuum, and then in air. The result of these measurements are presented and discussed in Sections \ref{sec:Vacuum} and \ref{sec:Ambient}, respectively.

\section{\label{sec:Vacuum}Heat transport under vacuum conditions}

\begin{figure}
	\includegraphics[width=8.5cm]{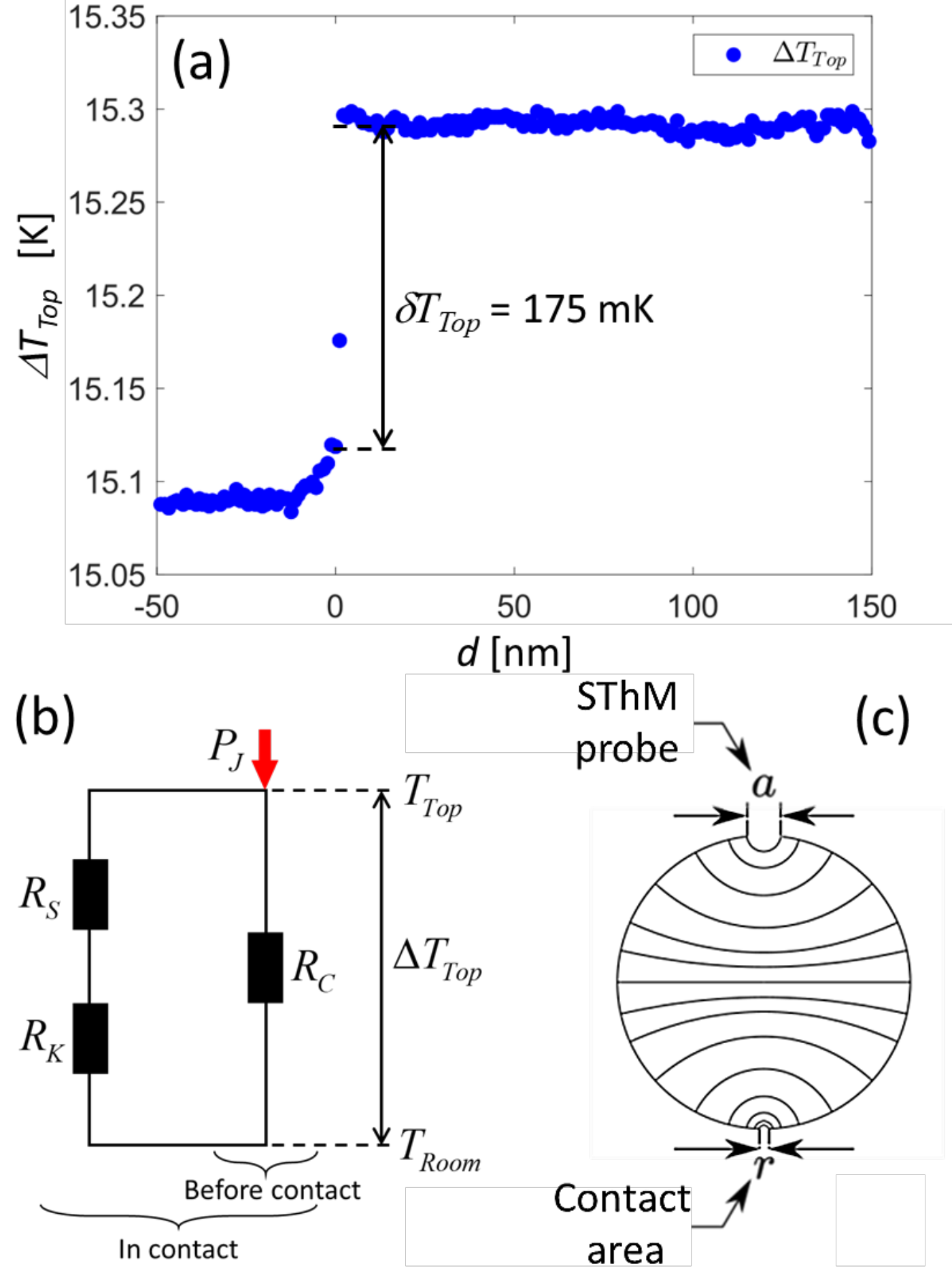}
	\caption{\label{figure_2}(a) Vacuum measurement with $P_J=71\mu$W of the temperature difference $\Delta T_{Top}=(T_{Top}-T_{Room})$ as a function of the distance $d$ between the base of the microsphere and the glass plate. $d=0$ corresponds to the occurence of the sphere-plate contact where a sudden temperature jump of amplitude $\delta T_{Top} =175$ mK occurs. (b) Scheme showing the various resistances through which heat can flow before and after the contact is established between the sphere and the plate. $R_C$ is the thermal cantilever resistance,    $R_S$   is the sphere resistance, and $R_K$ is the sphere-plate contact resistance. (c) Isotherms of an isolated sphere heated on a pole and cooled on the other. $a$ is the size of the palladium strip and $r$ is the equivalent size of contact.}
\end{figure}

The first measurements were performed in high vacuum at a pressure of $10^{-6}$mbar such that the mean free path of air molecules  $\approx 100$ m is much larger than the vacuum chamber.  Fig.~\ref{figure_2} (a) presents the results of temperature measurements recorded when the microsphere is approached towards the plate. In the experiment, a heat power $P_J=71\mu$W produces a temperature increase $\Delta T_{Top}=15.30$ K in the absence of contact between the microsphere and the plate. The total thermal conductance between the thermoresistive palladium strip and the environment before contact is thus $G_{Total}(d>0)= 4.6\ \mathrm{\mu W.K^{-1}}$.
The temperature at the top of the sphere as function of the distance is mostly constant and only exhibits an abrupt drop in amplitude $\delta T_{Top}\simeq175 \ $mK at $d=0$.
This drop corresponds to contact: a new conductive channel opens and substantially increases heat leakage. Note that besides the step-like variation, slow fluctuations are superimposed on the approach curve as the vacuum configuration requires the use of an electrical wiring which is not as well optimized as in air in terms of length, intermediate connections, and thermal shielding. 

The radiative conductance between a 20 $\mu$m diameter glass microsphere and a glass plate is only of a few $\mathrm{nW .K^{-1}}$ at maximum in the far field: $G_{ff}(T) = 2\pi R^24\sigma\varepsilon(T_{Sphere})T^3$, with $R$ the radius of the sphere, $\sigma$ the Stefan constant and $\varepsilon(T_{Sphere})$ the emissivity of silicon dioxide at the sphere temperature $T_{Sphere}$ \cite{palik_handbook_1997}.
For a sphere heated up to 45$\mathrm{^\circ}$C the conductance is $G_{ff}(T) = 1.6 \ \mathrm{nW.K^{-1}}$. This value increases by a few $\mathrm{nW.K^{-1}}$ in the near field. 
The radiative conductance has only been observed so far with SThM techniques provided that the temperature difference between the sphere and the plate was on the order of hundreds of Kelvin \cite{menges_thermal_2016,lucchesi_harnessing_nodate}. Its measurement with smaller temperature differences was possible with ultrasensitive fluxmeters \cite{rousseau_radiative_2009,cui_study_2017}. The sensitivity of our SThM based setup is deduced from the noise on the temperature measurement before contact which is found to be 10 mK, corresponding to a conductance sensitivity of 3 $\mathrm{nW .K^{-1}}$, and having the same order as the near-field contribution observed in \cite{rousseau_radiative_2009} for 22 $\mu$m glass sphere. Near-field effects are not observable at this stage with our setup for such a sphere diameter. 

Neglecting this contribution compared to conduction, the cantilever conductance can be directly inferred from measurements out of contact ($d>0$). Fig.~\ref{figure_2}(b) sketches the thermal resistance circuit to be considered in the experiment. It involves three thermal resistances associated with the cantilever, $R_C$, the sphere, $R_S$, and the sphere-plate contact, $R_K$, and the corresponding conductances $G_{C,S,K}=1/R_{C,S,K}$. Since before contact $R_K \simeq \infty$, the sphere is isothermal and all the heat dissipated at the palladium strip flows through the cantilever whose anchor point is  at room temperature. The cantilever conductance $G_C$ is equal to the total thermal conductance before contact: $G_{C}=G_{Total}(d>0)= 4.6\ \mathrm{\mu W.K^{-1}}$. 

Figure ~\ref{figure_2}(a) shows that the temperature measured at the top of the sphere drops abruptly by an amount $\delta T_{Top}$ at $d=0$ when the base of the sphere enters in contact with the plate. This drop corresponds to the opening of a new conduction channel for heat at contact, as illustrated in Fig.~\ref{figure_2}(b). Note that $\Delta T_{Top}$ still decreases by $\sim 25$ mK between 0 and -10 nm as the sphere is further pressed against the plate, which is likely due to a small deformation of the contact region. With the sphere in contact with the plate held at $T_{Room}$, heat can flow from the heated palladium strip through the sphere and the contact. The sphere is no longer isothermal. Its resistance is in series with the sphere-plate contact resistance, and the total conductance increases by an additional term $\left (1/G_{S}+1/G_{K} \right )^{-1}$ compared to its value $G_C$ before contact. 
\begin{equation}
G_{Total}(d=0)=\frac{P_J}{\Delta T_{Top}(0)}=\frac{1}{R_S+R_K}+G_C
\label{eq:Contact_resistance_equation}
\end{equation}	
where $\Delta T_{Top}(0)$ is the temperature difference at contact. This provides a way to calculate the contact resistance $R_K$ from the measured values of $ G_{Total}$ in contact and before contact.

The calculation of $R_{K}$ from the measurement of $G_{Total}(d=0)$ thus requires to determine the sphere thermal resistance $R_S$. The latter can be calculated in analogy to the electrical resistance of a sphere because the poles are considered as heat source and sink, and the surface of the sphere is isolated everywhere else due to a negligible radiative heat transfer. The problem of a conducting sphere was addressed analytically by Landau and Lifshitz \cite{Landau_1963} and Solivérez \cite{Soliverez_2012}. The resistance of a sphere of radius $R$ and conductivity $k$ whose poles are connected with equal-size contacts of radius $r \ll R $ is $R_{S}(r) = ({1}/{\pi k})\left[({1}/{r})+({1}/{2R})\left(\ln\left({4R}/{r}\right)-1\right)\right]  $
\cite{Soliverez_2012}. In our experiment, $k=(1\pm 0.1)\ \mathrm{W.m^{-1}.K^{-1}}$ is the conductivity of glass. Heat is injected via the palladium strip which forms a contact of radius $a=1 \ \mu$m with the sphere top, and then sinks via the sphere-plate contact of radius  $r$. We thus have two hemispheres with different contact radii in series. The sphere resistance is then $R_{S}(a,r)=\frac{1}{2}R_{S}(a)+\frac{1}{2}R_{S}(r)$,  which yields the expression of the resistance of a sphere of radius $R$ and conductivity $k$ with different contact radii $a$, $r$ at the north and south poles respectively: 

\begin{equation}
R_{S}(a,r) = \frac{1}{2\pi k}\left [\frac{1}{a}+\frac{1}{r}+\frac{1}{2R}\left(\ln\left(\frac{16R^2}{a\ r}\right)-2\right)\right] . 
\label{eq:Sphere_resistance_equation}
\end{equation}

According to Eq.(\ref{eq:Sphere_resistance_equation}), the sphere resistance is expected to strongly depend on $r$ which is a priori not known. However, since the phonon mean free path of glass near room temperature is less than 1 nm we assume that the transport of heat across the contact is purely diffusive \cite{Kittel_1949}. In this regime, the resistance is directly related to the constriction size: $R_K = {1}/({2 k r})$\cite{prasher_thermal_2006}. $r$ can thus be expressed in terms of $R_K$ in Eq. (\ref{eq:Sphere_resistance_equation}) and using Eq. (\ref{eq:Contact_resistance_equation}) and $G_{Total}(d>O)=G_C$ it is possible to determine both $R_K$ and $R_S$ from the measurement of the temperature jump at contact $\delta T_{Top}=\Delta T_{Top}(d>0)-\Delta T_{Top}(d\leq 0)$. 

The thermal resistance of the glass microsphere-plate contact in vacuum is obtained to be $ R_K=(1.4 \pm 0.18)
\times 10^7 \ \mathrm{K.W^{-1}}$, or equivalently $G_K=72 \ \mathrm{nW.K^{-1}}$. This corresponds to an effective contact radius $r=(36 \pm 4)$ nm. It follows that the sphere resistance is $R_S\sim 4.6\times 10^6 \ \mathrm{K.W^{-1}}$. With the values of the resistances found, we can thus calculate that only 1.2 $\%$ of the generated power flows through the sphere and the contact, while the majority of it flows through the cantilever. A series of measurements were performed for several temperature differences  ($\Delta T_{Top} = 2, 12, 15\ \mathrm{K}$) and at different locations of the glass plate. We find the same value of $ R_K\approx1.4 \ 10^7 \ \mathrm{K.W^{-1}}$. Note that a contact resistance of a few $10^7 \ \mathrm{K.W^{-1}}$ was also measured between two crossing glass micro-fibers \cite{Nguyen_JHT2020}. Let’s also mention that the contribution of the glue between the SThM probe and the sphere has been neglected in this analysis. Its thickness is $\approx$100 nm, which corresponds to a resistance of about $10^5 \ \mathrm{K.W^{-1}}$ given the conductivity $k_G\approx 0.3 \ {W.m^{-1}.K^{-1}}$ of the epoxy glue in use in the experiment.   

\section{\label{sec:Ambient}Heat transport under ambient conditions}
Similar experiments with a SThM probe from a different batch and a glass sphere with $D=20.7 \pm 1 \mathrm{\mu m}$ glued on its tip were also carried out under ambient conditions, i.e. in air under atmospheric pressure with a relative humidity of $(70\pm 10)\%$. These conditions are more complex to deal with since several heat loss channels coexist even when the sphere and the plate are out of contact. With separations between the sphere and the plate reaching up to hundreds of micrometers, heat can still be transferred through air whose thermal conductivity $k_a=(26.6 \pm 0.1)\times 10^{-3}\ \mathrm{W.m^{-1}.K^{-1}}$ is small but not enough to be neglected. Other heat transfer channels may also contribute as both the sphere and cantilever constitute fins. While the thermal problem could probably be treated exactly by solving the heat equation with appropriate boundary conditions, we adopt in what follows a phenomenological model based on thermal resistance circuitry to describe the temperature variations recorded at the top of the sphere when it approaches the plate from hundreds of micrometers to contact. This procedure is sufficient to obtain a good estimate of the thermal-contact resistance between the sphere and the plate in air. 

\begin{figure}
	\includegraphics[width=8.5cm]{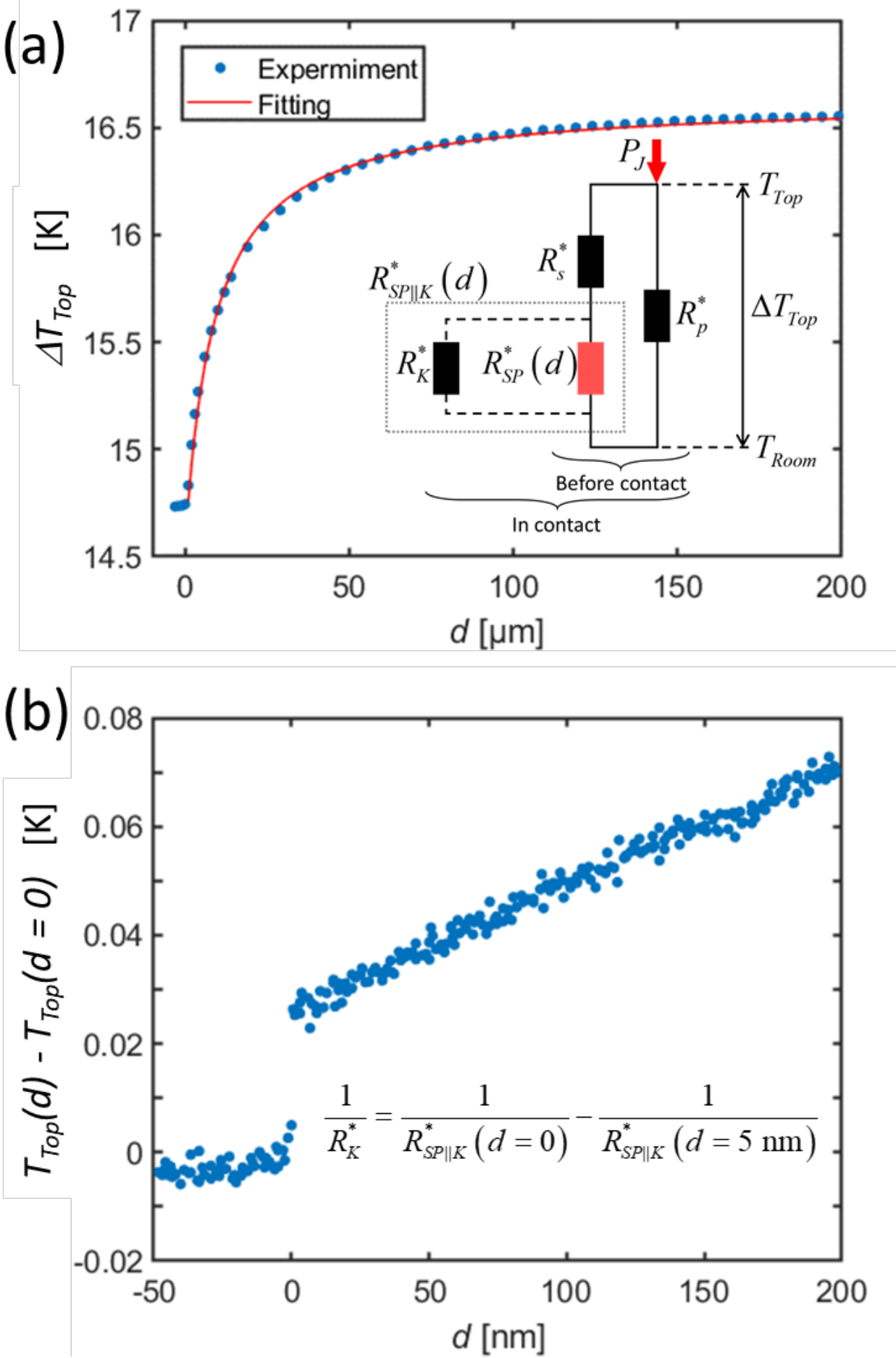}
	\caption{\label{figure_Air}(a) (blue dots) Experimental approach curve showing the temperature of the palladium sensor glued on the north pole of the sphere as a function of the distance $d$ between the south pole and the plate. $d=0$ corresponds to the sphere-plate contact. The red curve is the result of a fit based on the conduction shape factor and the resistance circuit shown in the inset. (b) Approach curve at short distance showing the jump of temperature occurring at contact due to the opening of a new conduction channel.}
	\end{figure}

Figure ~\ref{figure_Air} (a) shows the variation of $\Delta T_{Top}$ as a function of the sphere-plate separation $d$ recorded in air from 200 $\mu$m to contact with $P_J=68.4\ \mathrm{\mu W}$. High accuracy measurements of the temperature drop occurring at contact are presented in Fig.~\ref{figure_Air} (b). At  large distances (Fig.~\ref{figure_Air} (a)) , the curve first exhibits a slow decay as $d$ is reduced, which increases below 100 $\mu$m. As in vacuum, a jump in temperature is observed at contact, but with a reduced amplitude of $(28 \pm 3) \ $mK.    

The underlying idea of the model is to separate the thermal resistance due to the sphere-plate coupling from other heat-loss channels which do not depend on the distance $d$.  When approaching the sphere to the plate in air, the contact at $d=0$ produces a small additional contribution to the sphere-plate coupling by conduction through air $R^*_{SP}(d)$, whose thermal resistance $R^*_K$ can be evaluated by measuring the temperature jump. The thermal resistance circuit on which the model relies is sketched in the inset of Fig.~\ref{figure_Air}(a). We approximate the various heat-loss channels by a set of three thermal resistances: a series resistance $R^*_s$ and a parallel resistance $R^*_p$, both representing the distance independent losses, and a resistance  
$R^*_{SP||K}(d)=R^*_{SP}(d)||R^*_K=(1/R^*_{SP}(d)+1/R^*_K)^{-1}$ which describes the loss channel between the sphere and the plate. The latter includes two resistances placed in parallel: $R^*_{SP}(d)$ which depends on  $d$ and $R^*_K$ the contact resistance which takes a finite value only at contact ($d\leq 0$). Since $R^*_K=\infty$ out of contact, we have $R^*_{SP||K}(d)=R^*_{SP}(d)$ for $d>0$.

In contrast with the vacuum case (section ~\ref{sec:Vacuum}), a fitting procedure is needed to find the values of $R^*_s$ and $R^*_p$ . Regarding the sphere-plate coupling resistance out of contact $R^*_{SP}(d)$, we found that the conduction shape factor \cite{Kreith} of an isothermal sphere placed close to a semi-infinite isothermal plane $S(d)= ({2 \pi D})/\big[{1- {D}/{4 \big(d+D/2\big) } }\big] $ describes  the coupling very well. This  suggests that a large portion of the glass sphere is quasi-isothermal. We thus set $R^*_{SP}(d)={1}/\big({S(d) \  k_a}\big)$ in the fit. In Fig. ~\ref{figure_Air} (a) we show the fit of the approach curve obtained in this way, superimposed on the experimental curve. The fit corresponds to $R^*_s=(2.84 \pm 0.13)\times 10^5 \ \mathrm{K.W^{-1}}$ and $R^*_p=(4.22 \pm 0.08)\times 10^5 \ \mathrm{K.W^{-1}}$. Note that these resistances are the only  fitting parameters of the model. They mimic the losses through the solid phases as well as through air, provided that these losses are independent of the sphere-plate distance.  
Once the values of $R^*_s$ and $R^*_p$ have been determined, we use the experimental values of $\Delta T_{Top}$ measured at $d=5$ nm and $d=0$ (contact), i.e. on both sides of the jump in Fig. ~\ref{figure_Air}(b), to calculate the corresponding values of $R^*_{SP||K}(d)$. The occurrence of the sphere-plate contact coincides with the opening of a new conduction channel besides the coupling through air. We thus have 
%$\frac{1}{R^*_K}=\frac{1}{R^*_{SP}//R^*_K\big(d=0\big)}-\frac{1}{R^*_{SP}//R^*_K\big(d= 5 nm\big)}$,
${1}/{R^*_K}={1}/{R^*_{SP||K}(d=0)}-{1}/{R^*_{SP||K}(d=5 nm)}$,
which yields the value of the contact resistance in air $R^*_K=(1.2 \pm 0.46)\times 10^7 \ \mathrm{K.W^{-1}}$. This value is close to that found in vacuum, which suggests that if a water meniscus is present within the contact area between the sphere and the plate, it only has a minor influence on the value of the thermal contact resistance. This is consistent with reported results obtained with SThM probes \cite{assy_temperature-dependent_2015}. Finally,
the resistance circuit shown in the inset of Figure \ref{figure_Air} (a) enables one to calculate the relation which links $R^*_K$ and  $\Delta  T_{Rel}=\delta T_{Top}/\Delta T_{Top}(d=0)$, the temperature jump occurring at contact measured at the top of the sphere relative to the temperature difference with respect to $T_{Room}$. $
R^*_K =\left (\Delta  T_{Rel}^{-1}\right )\left (  \left ( R^{*2}_{SP} R^{*}_{p} \right )  \left ( R^{*}_{s}+ R^{*}_{SP}\right )^{-1}  \left ( R^{*}_{s}+ R^{*}_{SP}+ R^{*}_{p} \right )^{-1}\right )$. The minimum jump of temperature which can be resolved by our SThM setup in air is $\approx$ 5 mK around $\Delta T_{Top}$ =14.5 K, i.e.  $\Delta  T_{Rel}=3.4 \ 10^{-4} \ \mathrm{K.K^{-1}}$. The highest value of $R^*_K$ which could thus be measured is $\approx 6.8 \ 10^7 \ \mathrm{K.W^{-1}}$ in air at atmospheric pressure, while we estimate it to be  $\approx 3.3 \ 10^8 \ \mathrm{K.W^{-1}}$ in vacuum. An improvement of the sensitivity by one order of magnitude should still be possible by modifying the measurement electronics to reduce the uncertainty on $\Delta T_{Top}$ below 1 mK.

\section{\label{sec:Conclusion}Conclusion}
A quantitative characterization method of the thermal resistance between a glass microsphere and a glass plate in vacuum or in air has been demonstrated. Using a SThM tip glued on the sphere as both a heater and a temperature sensor, it is possible to vary the sphere-plate separation with nanometer precision from hundreds of micrometers to contact. Both in vacuum and air, a jump of temperature is observed at contact due to the opening of a new conduction channel. In spite of the low thermal conductivity of glass and of the complex geometry of the experiment (the contact is opposite to the temperature sensor on the glass sphere), it was possible to disentangle the value of the thermal contact resistance from the height of the jump.  In vacuum, we could measure the thermal resistance of the SThM cantilever in use. With an analytical expression of the thermal resistance of a sphere with different contact radii at the poles, and assuming diffusive heat transport, we were able to determine the value of the sphere-plate thermal contact resistance and the contact size, which we found to be of tens of nanometers. In air, the measurements showed that the distance dependence of the approach curve is governed by conduction through the gas phase. Based on a phenomenological model involving the conduction shape factor and a fitting procedure it was then possible to estimate the sphere-plate thermal contact resistance in air. We found a value close to that  measured in vacuum. 

The proposed methods have the advantage of being very generic and based on analytical modelling. With little effort, they could be adapted to estimate the thermal contact resistance between micrometer sized objects made of materials other than glass or with different shapes, or to study the impact of the environment on the transport of heat at small scales. This opens large perspectives for future  investigations of heat transfer processes at a microscopic scale between the elements of complex insulation materials used in the building sector. For instance, to determine the contact resistance between glass fibers crossing with an arbitrary angle like in glass wool, it should be possible to calculate numerically only a few points of the curve $R^*_{SP}(d)$ in the simplified geometry of two crossing cylinders without  cantilever, and to determine $R^*_s$ and $R^*_p$ by a fit of the experimental data, which would provide enough parameters to calculate $R^*_K$. 

\appendix*
\section{\label{sec:Appendix}Thermistance measurement instrument }
\begin{figure}[b]
	\includegraphics[width=8.5cm]{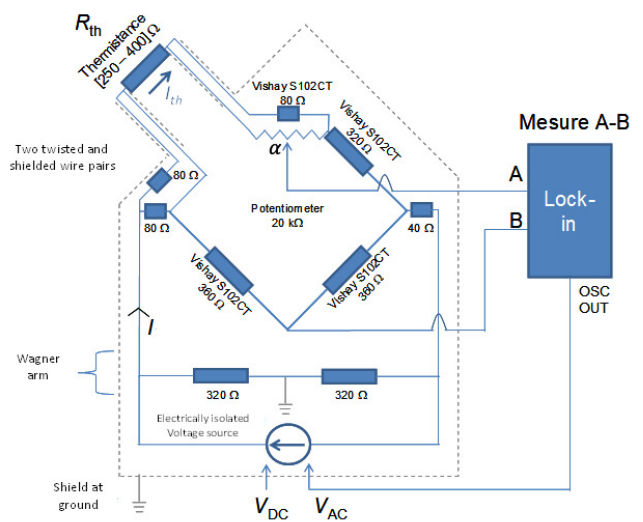}
	\caption{Scheme of thermistance bridge}
	\label{figure_Appendix}
\end{figure}

The thermal sensor of the SThM probe is based on a palladium resistive component with a nominal resistance of about 320 $\Omega$ at room temperature and a typical temperature dependence of 0.7 $\Omega /K$. The measurement of the temperature therefore consists of measuring a resistance value which varies as a function of the temperature with a law that can be calibrated. Extracting quantitative information about the thermal exchanges in the experiment requires the knowledge not only of the temperature but also of the joule power which can be set by supplying the resistance sensor with a known current. To achieve this double objective, the measurement instrument, described in fig.~\ref{figure_Appendix}, is based on a Wheatstone bridge allowing the combination of an AC (alternating current) measurement of the temperature and of a DC (direct current) heating of the sensor. The resistance value, $\mathcal{R}$, of the thermal sensor (thermistance) is obtained from the three other reference resistors and the voltage unbalance of the bridge according to the relationship:

\begin{equation}
\mathcal{R}\approx \frac{360\times \left ( 400-160\alpha  \right )-377600\gamma }{360+800\gamma }
\label{eq:Bridge_Equation}
\end{equation}
where $\gamma =\frac{V_{AB}}{V_{AC}}$, $V_{AB}$ is the amplitude of the ac unbalance voltage measured by the lock-in detector, $V_{AC}$ is the ac applied voltage to the Wheatstone bridge and $\alpha \in \left [0..1  \right ]$ is the potentiometer setting. The DC and AC currents,  $I_ {DC}$ and $I_ {AC}$ that circulate through the thermal sensor, are given by:
\begin{equation}
I_{DC,AC}\approx \frac{720V_{DC,AC}}{377600+800\mathcal{R}}
\label{eq:Current_Bridge_Equation}
\end{equation}

The setting $\alpha$ of the 20 k$\Omega$ potentiometer in parallel with the 80 $\Omega$ resistor is used to balance the bridge ($V_{AB}\approx 0$) in order to accommodate for sensor resistance values away from 320 ohms. Detection of a low voltage signal allows choosing the highest sensitivity range of the lock-in detector. The Wheatstone bridge being equipped with a Wagner arm with a mid-point at ground, permits both $V_A$ and $V_B$ detection voltages to be close to the ground potential. This allows the absence of a common-mode voltage which is favorable for having better sensitivity and accuracy. 

To obtain good reproducibility and stability of measurements, several techniques were implemented \cite{Schopfer_JAP2007,Schopfer_JAP2013,Poirier_IEEE2020}. First, the bridge is made of S102CT Vishay resistors having a relative drift below $10^{-5}$/year and a low temperature coefficient below 0.6 ppm/K. The electronic circuits delivering the biasing AC and DC voltages are based on precision operational amplifiers supplied by stabilized voltage sources powered by rechargeable batteries. They are strongly isolated from the ground with PTFE material. Second, the thermal sensor is connected with 4 wires in a way (a triangle of two 80 $\Omega$ resistors on one side and one 80 $\Omega$ resistor and the potentiometer on the other side) that makes the bridge balance more immune to any change (for example caused by temperature) of the wire resistances. This implementation is nevertheless fully effective for a thermal sensor resistance close to 320 $\mathrm{\Omega}$. Finally, pairs of high and low potential wires are independently twisted and placed in a shield at ground as the case of the bridge instrument. Because of the mid-point at ground of the Wagner arm, current leakages between wires due to capacitance and insulating resistance are redirected towards the ground without changing the balance of the bridge. This avoids measurement errors, and circulation of high-frequency electrical noise, and allows increasing the measurement bandwidth. The efficiency of the electric guarding system is also optimum for $\mathcal{R}\approx 320 \ \Omega$. 

Close to equilibrium, i.e. $\alpha$ is set so that $\gamma\approx 0$, the resistance sensitivity is given by:
\begin{equation}
\delta \mathcal{R}\approx -\frac{\left ( \frac{377600}{316800} \right )\delta V_{AB}}{I_{AC}}\approx -1.2\times  \delta V_{AB}/I_{AC}
\label{eq:Sensitivity_Bridge_Equation}
\end{equation}
Considering a voltage detection sensitivity of 5 $\mathrm{nV/{Hz}^{1/2}}$ and an AC polarization current $I_{AC}$ of 20 $\mu$A, a resistance sensitivity of $3\times 10^{-4}\Omega /\rm{Hz}^{1/2}$ can be achieved which corresponds to a temperature sensitivity of about 1 mK/$\rm{Hz}^{1/2}$ at best.

\begin{acknowledgments}
	This research was supported by LABEX WIFI (Laboratory of Excellence within the French Program Investments for the Future) under references ANR-10-LABX-24 and ANR-10-IDEX-0001-02 PSL*, and Agence Nationale de la Recherche (ANR), Project CarISOVERRE, ANR-16-CE09-0012. The authors thank K. Joulain, H. Kallel (Inst. Pprime), O. Bourgeois (Inst. Néel), and J.-Y. Laluet, V. Grigorova-Moutiers, J. Meulemans (Saint-Gobain) for their constructive comments and encouraging support. 
\end{acknowledgments}

\bibliography{Biblio_PRApplied_Doumouro}

\end{document}